\documentclass[preprint]{aastex61}
\begin{document}

\title{Followup Observations of SDSS and CRTS Candidate Cataclysmic 
Variables II.\footnote{Based on observations obtained with the Apache Point
  Observatory (APO) 3.5-meter telescope, which is owned and operated
  by the Astrophysical Research Consortium (ARC).}}

\author{Paula Szkody}
\affiliation{Department of Astronomy, University of Washington
  Box 351580, Seattle, WA 98195; szkody@astro.washington.edu}

\author{Mark E. Everett}
\affiliation{National Optical Astronomy Observatories, 950 N. Cherry Ave, 
Tucson, AZ}

\author{Zhibin Dai}
\affiliation{Department of Astronomy, University of Washington,
  Box 351580, Seattle, WA 98195}
\affiliation{Key Laboratory for the Structure
and Evolution of Celestial Objects, Yunnan Observatories, Chinese Academy of
Sciences, 396 Yangfangwang, Guandu District, Kunming, 650216, P. R. China}
\affiliation{Center for Astronomical Mega-Science, Chinese Academy of
Sciences, 20A Datun Road, Chaoyang District, Beijing, 100012, P. R. China}

\author{Donald Serna-Grey}
\affiliation{Department of Astronomy, University of Washington,
  Box 351580, Seattle, WA 98195}

\begin{abstract}
Spectra of 38 candidate or known cataclysmic variables are presented. Most are 
candidate
dwarf novae or systems containing possible highly magnetic white dwarfs, while
a few (KR Aur, LS Peg, V380 Oph amd V694 Mon) are previously known objects 
caught in unusual states. Individual spectra 
are used to confirm a dwarf nova nature or other classification while radial 
velocities of 15 systems provide orbital periods and velocity amplitudes that
aid in determining the nature of the objects. Our results substantiate a
polar nature for four objects, find an eclipsing SW Sex star below the
period gap, another as a likely intermediate polar, as well as two dwarf
novae with periods in the middle of the gap.
\end{abstract}

\section{Introduction}

This is the second paper in a series containing follow-up observations of
candidate cataclysmic variables (CVs) 
found in various past and ongoing sky surveys
such as the Sloan Digital Sky Survey (SDSS; \citet{Y00}), the Catalina
Real-Time Transient Survey (CRTS; \citet{D09}), the All Sky
Automated Survey (ASAS; \citet{P97} and ASAS-SN; \citet{Sh14}) and the Mobile Astronomical System
of the Telesope Robots (MASTER; \citet{L10}). As described in
Paper I \citep{S14}, the confirmation and properties of the candidates
found in the photometric surveys require spectra and further orbital light
curves. Most candidate objects are dwarf novae, which are easily discovered in
the sky surveys since
they undergo periodic outbursts due to a disk instability resulting from
the accumulation of the mass transferred from the late
type companion onto an accretion disk that ultimately accretes onto
the primary white dwarf star. Since the periods of the outbursts depend on
the mass transfer rate, the highest transfer rate objects are preferentially
found while the shortest orbital period systems, with outburst timescales
of decades, can remain hidden until an outburst occurs. Thus, significant
biases exist in the determination of the real CV population. Hidden among
 these CV systems are those that contain highly magnetic white
dwarfs, the polars and intermediate polars (different types are reviewed
in Warner 1995) and other novalikes which undergo high and low states when
the mass transfer is either on or off (SW Sex, VY Scl subtypes). These latter 
systems tend to have
high excitation lines of \ion{He}{2} present when they are active, thus 
requiring
spectra to confirm their identity. Long term photometry can suggest their
nature due to the large amplitude orbital modulations that are present on
orbital timescales of hours due to the different viewing perspectives of
the magnetic accretion columns during the orbit.

Paper I presented spectra of 35 systems obtained from 2010 September-2013 
October, while this paper contains additional spectral data on 38 systems
from 2014 March-2017 March.
When combined with spectra from other groups \citep{T12,B14,T16},
 these data allow future 
global
studies of a confirmed population of objects. For simplicity, we provide
identification of all objects by their 2000 coordinates in Table 1
 (allowing them to
be found in their photometric databases), while we abbreviate those coordinate 
in the following sections.
 
\section{Observations}
The majority of the spectra were obtained with the Double Imaging Spectrograph
on the Apache Point Observatory (APO) 3.5m telescope. The high resolution
(0.6\AA\ pixel$^{-1}$) gratings were used to simultaneously 
2015 September only the blue spectrograph was available). 
Flux standards and HeNeAr lamps were
used to provide calibrated spectra and spectra were reduced using 
IRAF\footnote{IRAF is distributed by the National Optical Astronomy Observatory,
which is operated by the Association of Universities for Research in Astronomy,
under cooperative agreement with the National Science Foundation.} standard
routines.

A few spectra in 2014 were obtained at the Kitt Peak
National Observatory (KPNO) 4m telescope with the RC Spectrograph,
using grating KPC-22b in second order and a 1 arcsec slit to produce blue
spectra from 3800-4900\AA\ with a resolution of 0.7\AA\ pixel$^{-1}$.
FeAr lamps and flux standards were used along with IRAF reductions to
produce final calibrated spectra.

Table 1 summarizes all the spectra obtained while a sample blue spectrum of
each object is shown in Figure 1. Table 2 shows the equivalent widths of
the emission lines of H$\alpha$, H$\beta$ and \ion{He}{2}4686 as measured with
the $\it{e}$ routine in the IRAF $\it{splot}$ package.

\section{Results on Systems with Time-Resolved Spectra}

For 18 of the systems in Table 1, five or more time-resolved spectra were
obtained. The strongest Balmer lines (usually H$\alpha$ and H$\beta$ along
with \ion{He}{2}4686 in a few cases)
were used to determine the
centroids and compute velocities using the $\it{e}$ routine in the IRAF 
$\it{splot}$ package. For 0116+09 with strongly doubled lines,
the Double-Gaussian method \citep{S83} was also
used. The velocities were then fed into
software programs to compute the best least-squares fit to a sinusoid, yielding
$\gamma$, semi-amplitude K, Period and the total $\sigma$ of the fit.
Results were obtained for 15 of the systems, with fit parameters listed
in Table 3, and the individual results are described below. Examples of the 
radial velocity curves (for the strongest line with the best fit), are shown
in Figure 2. Among the remaining three systems, 1545+01
(ASSASN-14cm) was at outburst so the broad absorption lines did not reveal
the period. The weak lines in 0929+62 did not show any velocity change
$>$ 31 km s$^{-1}$ during the 80 min of spectra. While the spectra of
2200+03 appear similar to dwarf novae, a satisfactory solution
could not be obtained (likely due to insufficient
time coverage during the 97 min of observation for what appears from the
velocities to be a
long orbital period of $\sim$160-200 min). 

\subsection{V677 And: likely polar}

V677 And was reported as a flaring transient in CRTS as CSS080924:233423+391423
by \citet{M08} and spectroscopically identified as a likely dwarf
nova by \citet{Q08}. Later photometry by \citet{C15} revealed a
period near 100 min along with speculation that it could be a polar. Our
12 spectra obtained throughout 3 hrs showed strong Balmer emission lines
in the blue, along with \ion{He}{2}4686 line comparable to the strength of 
H$\beta$
 (Figure 1b).
A large periodic velocity variation was evident throughout the time series
and fits to a sine wave yielded periods of 105 min and very large K 
semi-amplitudes of 400 km s$^{-1}$ (Table 3, Figure 2). The short period, high
K amplitude, combined with the 1.2 mag variations evident in photometry
make it very likely that V677 And is a polar containing a highly magnetic
white dwarf.

\subsection{V380 Oph: SW Sex in low state}

This object was determined to be a novalike cataclysmic variable in a
spectroscopic survey by \citet{B79}, and studied in more detail by 
\citet{S85}
who found a period of 3.8 hr and a radial velocity semi-amplitude of 100$\pm$14
km s$^{-1}$ when the system was at a magnitude of about 15.5. \citet{S05}
searched plate files and determined high and low states existed
in a range from 14.5-17 mag. \citet{R07} accomplished a
more thorough study at higher spectral resolution, improving the period
to 3.69857 hr (221.9 min) and finding a K semi-amplitude of 207$\pm$5 km 
s$^{-1}$ when the system was at a comparable magnitude to the time when Shafter
had observed. They concluded that V380 Oph was a high mass transfer SW Sex
type novalike\footnote{See D.~W.\ Hoard's Big List of SW Sextantis Stars at 
\url{http://www.dwhoard.com/biglist}
\citet{H03} for a complete discussion and current list of SW Sex stars} and could be an IP, as they found a possible 
47 min periodicity
in their data which could be the spin period of a magnetically accreting
white dwarf. In 2015 July, V380 Oph began a low state near 17th mag
and we obtained spectral coverage of 0.76 of its orbit. Comparing our
low state spectra (Figure 1b) to those of \citet{S85} and \citet{R07} 
during high states shows a much flatter blue continuum that
has almost a factor of 10 less flux than the published high state spectra.
The equivalent widths of our Balmer emission lines of H$\beta$
of 22 \AA\  (Table 2) are more than double those during the high
state. Surprisingly, our K amplitude of 71$\pm$3 (Table 3, Figure 2) 
is even lower than the
value found by \citet{S85} and almost 3 times lower than the value of
\citet{R07}. While determining masses from emission lines
is unreliable, the differences between low and high states are consistent
with a shrinking disk, allowing measurment of a more massive white dwarf
than can be viewed during a high state. 

\subsection{0038+25: DN}

This g=18.85 blue object found in Data Release 12 of the SDSS shows
a steep blue continuum with strong doubled Balmer and HeI lines as well as TiO
band features from a red companion in the SDSS spectrum (Figure 1a). Our 10 blue spectra
covering almost 2 hr reveal a typical short period CV (95 min) with a
K value of 95$\pm$13 km s$^{-1}$ (Table 3, Figure 2).

\subsection{0116+09: DN}

The CRTS discovered an outburst at 16 mag on 2008 Dec 20, while the SDSS g mag 
is 19.1. The long term light curve shows several outbursts over a 10 year span.
Our quiescent spectra show the broad doubled Balmer lines typical of a
dwarf nova (Figure 1a). 
The doubled lines corroborate the photometric result of 
\citet{C14} that 0116+09 is a deeply eclipsing CV system. 
 Based on our 19 spectra at quiescence observed on 2014
September 2 with a spectral coverage of 2.45 orbits, the two trailed spectra of
H$\alpha$ and H$\beta$ (Figure 3) clearly show evidence of an S-wave, which 
indicates 
the motion of a hot spot on the disk. Since the H$\beta$ emission line is 
always deeply doubled throughout the whole orbit, the brighter H$\alpha$ 
emission line was used to derive its orbital period. Using the 
Double-Gaussian method developed by \citet{S83}, our 19 quiescent spectra 
provide a radial velocity curve from H$\alpha$ revealing a period of 
93.6$\pm$1.3\,min and a K semi-amplitude of 75.6$\pm$1.3\,km\,s$^{-1}$.  
Using the $e$ routine in IRAF to determine the centroids of H$\alpha$ and 
H$\beta$, the derived velocity curves derive similar periods of 91.4\,min 
and 87.9\,min, respectively. Although the K semi-amplitude for H$\alpha$ 
from the Double-Gaussian method is significantly smaller than that derived from
the $e$ routine (94.1$\pm$4.5\,km\,s$^{-1}$), it is consistent with the K 
semi-amplitude for H$\beta$ (74.9$\pm$0.3\,km\,s$^{-1}$) derived from
 the $e$ rountine.  The numbers are typical for a quiescent dwarf nova,
which further corroborates its dwarf nova identification.

\subsection{0333+33: likely polar}

The CRTS first revealed an increased brightness of this object in 2011 November.The light curve reveals two states, a high one at about 17.5 and a low one
about 20.5 with each state lasting years. During 5 nights of photometry in 2015,
\citet{L15} determined an orbital period of 110.53 min, with
the light curve showing large periodic variations from V=18 to 20.7 with
double humps per orbit as well as a dip lasting for 0.4 of the orbit. They
postulated a polar nature as the cause of the large variations. Our
spectra obtained in 2015 December corroborate this classification. The
narrow Balmer lines (Figure 1a), highly variable strength of \ion{He}{2}
throughout an orbit, and the very high K amplitudes (220-240 km s$^{-1}$) of the
H$\alpha$ and H$\beta$ lines (Table 3, Figure 2) are all consistent with polar spectral
characteristics.

\subsection{0411+23: DN}

This object is listed in the CRTS archive as showing a range in magnitude
from 15.5 to 18.5 with the comment of "not blue". The KPNO and APO spectra
 (Figure 1a) show strong but fairly narrow Balmer emission lines. The 138 min
of time-resolved spectra reveal a long period near 200 min with a low
velocity amplitude (Table 3, Figure 2). These characteristics explain the color comment as
well as the narrow lines if the inclination is low. Further data are needed
to obtain a more precise period.

\subsection{0501+20: DN}

The CRTS discovered this dwarf nova when it went into outburst in 2009 October.
The long term light curve shows a quiescent magnitude near 17.8 and many
outbursts with amplitude about 3 mags. Our quiescent spectra taken over a
2 hr interval show typical broad Balmer emission lines (Figure 1a) while the
radial velocities indicate a short orbital period of 108 minutes and
typical K semi-amplitudes of 84$\pm$5 and 51$\pm$19 for the H$\alpha$ and
H$\beta$ lines (Table 3, Figure 2).

\subsection{0648+06: DN in Period Gap} 

A large brightness change from 18 to 11.5 mag 
reported in the vsnet-alerts in 2014 November 22
 \citep{S14} resulted in a possible nova variable (PNV) designation
for this object. A spectrum by \citet{M14} near the same time 
was more consistent with a dwarf nova at outburst than a nova. Our spectra
3 weeks later (Figure 1a) confirm this identification, with strong, broad Balmer
emission lines typical of a quiescent dwarf nova. Our 2.5 hr of spectral
coverage reveal a period of 2.4 hr, squarely in the middle of the period
gap in the orbital period distribution of CVs \citep{W95}. The K 
semi-amplitudes of 70 km s$^{-1}$ are typical for dwarf novae (Table 3, Figure 2).

\subsection{0853+48: DN in Period Gap}

This CV found in SDSS DR12 also reveals a period in the middle of the
gap (142 min). The SDSS g magnitude is 18.9, and the SDSS spectrum shows strong
Balmer and HeI emission as well as a TiO and red flux from its late
type companion. Our time-resolved spectra show the Balmer emission lines
are broad and at times very doubled in appearance (Figure 1a). This, together
with K amplitudes near 100 km s$^{-1}$, imply a higher inclination than
for 0648+06 (Table 3, Figure 2).

\subsection{1005+69: likely IP}

The first spectrum of 1005+69 was obtained by the SDSS and suggested a
possible IP nature due to the presence of strong \ion{He}{2} \citep{S11}.
We accomplished three time-resolved spectroscopic observations in 2014 March 23,
May 4 and 6. The details are listed in Table 1. All spectra show 
strong single-peaked and asymmetrical Balmer emission lines in the red and blue,
along with strong \ion{He}{2} comparable to the strength of H$\beta$ and 
several 
weak HeI emission lines (Figure 1a). The 5 spectra covering almost 1\,hr obtained in the 
first two observations show a velocity variation larger than 100\,km\,s$^{-1}$.
This large velocity change was further confirmed by the following 13 spectra 
covering over 3\,hr obtained on 2014 May 6. The first five Balmer 
emission lines and the \ion{He}{2} line are all strong enough for the 
measurement of 
radial velocities, and the resulting velocity curves indicate a consistent 
orbital period of 220\,min and high K semi-amplitudes of 210-250\,km\,s$^{-1}$ 
(Table 3). Although the derived orbital period is typical for an SW Sex star 
(3-4\,hrs) and all the blue spectra show a relatively flat continuum flux like 
the SW Sex star V380\,Oph in a low state, the large radial velocity 
semi-amplitude 
of 211\,km\,s$^{-1}$ for H$\alpha$ is typical for a CV with a magnetic white dwarf.. 
The phases of the H$\alpha$ and H$\beta$ velocity curves are nearly the 
same as that of \ion{He}{2}. Moreover, the common spectral feature of an 
SW\,Sex star 
of a central absorption in the Balmer lines and HeI4471 near phase 0.3-0.5 is 
not evident in the layout of all the blue spectra (Figure 4). Thus, we
postulate 1005+69 to be an IP with a moderately magnetic white dwarf. The 
finding of a spin period of the white dwarf in further better time-resolved 
photometry can corroborate its plausible IP identification.

\subsection{1245-07: DN}

This object was detected at about 14.5 mag in 2015 May by both CRTS and
ASAS-SN (as ASASSN-15iq). It exists in the SDSS archive with a g mag of 19.8 and very
blue $u-g$ color of -0.05. The CRTS lists it as a possible polar. Our APO
spectra (Figure 1a) disputes this possibility as the lines clearly show
doubling from an accretion disk. Our 109 min of spectral coverage indicate
a period near 118 min and a K value of 60 km s$^{-1}$ (Table 3, Figure 2), 
which are typical
for a dwarf nova.

\subsection{1432+19: eclipsing SW Sex}

By using an outlier-mining method to pick up unusual objects in the
SDSS DR8, \citet{W13} found this 18.4 mag CV. The SDSS spectrum shows
a very blue continuum and \ion{He}{2} much stronger than H$\beta$. While our
spectra (Figure 1b) show a similar enhancement of \ion{He}{2}, the continuum
flux is about a factor of 3 larger. Our time resolved spectra over 2 hrs
show several changes in the line shapes, fluxes and velocities. The radial
velocity curve (Figure 2) shows large red and blue deviations from a sine
wave in 2 consecutive 10 min spectra, indicative of an eclipse (the 
Rossiter-McLaughlin effect). The continuum
drops during these 2 spectra, corroborating an eclipse. The radial velocity
curve for \ion{He}{2} shows the red to blue crossing at this time as well.
While the H$\beta$ velocity curve is noisier, the phases are offset by
about +0.1 from the \ion{He}{2} one. Furthermore, at phases 0.1-0.3, the
Balmer and HeI4471 lines show a central absorption, while they are 
single-peaked at other phases. All of these characteristics are the signature
of an SW Sex star. However, the period determined is much shorter than
the usual SW Sex systems which are between 3-4 hrs.\footnote{\url{http://www.dwhoard.com/biglist}} There are only 2 objects with orbital periods
below 2 hrs (EX Hya and SDSS J210131.26+105251.5) and these two are merely
listed as possible, not definite. Since our spectra only cover 115 min
and the eclipse removes two of these from a solution, further spectra
and photometry will be needed to determine a better period. If this short
period is determined to be real, then this could be a very interesting
test case for SW Sex accretion phenomena.

\subsection{2112-06: likely polar}

ASAS-SN reported this object on 2016 Oct 27 as ASASSN-16me, a CV candidate
with an SDSS g mag=19.7 and an unusual CRTS light curve. \citet{L16} 
obtained time-resolved photometry and a low resolution spectrum and
classified it as a deeply eclipsing polar with a period of 95.7 min. They
commented that their single spectrum showed surprisingly weak \ion{He}{2} but
did not know if this was just due to the orbital phase. Our 
spectra (Figure 1b) which cover 0.84 of an orbit show a similar weak flux for 
this high excitation line, with the line actually disappearing for part
of the orbit. Our radial velocity curve finds a very high K value (359 km 
s$^{-1}$) for H$\alpha$ (Table 3, Figure 2), consistent with a polar system.

\subsection{2319+33: DN}

This object is listed in the CRTS CV candidate list of \citet{D14}.
Its light curve shows outbursts to about 16.4 mag, and possible low states to
18.7. The SDSS lists $g$=17.79 and a very blue color of $u-g$=-0.34.
The two hours of spectroscopic coverage provide about 0.74 of the
orbital period determined from the radial velocities to be 171 min. This
period places it just inside the upper edge of the period gap. The
broad Balmer emission lines (Figure 1b) and average K value of 55 km$^{-1}$
(Table 3, Figure 2) are typical for a dwarf nova.

\subsection{2350+28: DN near P Mininum}

This Gaia discovered object (Gaia14ade) was confirmed as a CV with a low
resolution spectrum by \citet{W14}. While our best five spectra
in 2015 December only cover about 85 min and the Balmer emission is weak
(Figure 1b), the H$\alpha$ velocities undergo a smooth sinusoid during this 
time (Figure 2) yielding a very short period of 77 min, which is near
the observational period minimum for dwarf novae \citep{G09}. The K amplitude is typical for a dwarf
nova.

\section{Spectra at Outburst}

Spectra of four of the systems shown in Figure 1 were obtained at
outburst (0626+24, 1545+01, 1647+62 and V694 Mon). The first three show
the typical Balmer absorption lines that are evidence of a dominant
thick accretion disk near outburst peak. The last system is the symbiotic
star (previously known as MWC560) which had a previous outburst in 1990.
\citet{M16} summarize their photometric observations of the
2016 outburst as well as a long term lightcurve from 1928-2016. They found
that the 2016 outburst was about 0.2 mag brighter in B than the 1990 one. The 
2016 outburst had two peaks (Feb 7 and April 3) at B=9.2, so the spectrum in 
Figure 1a was obtained midway between these peaks. Broad P Cygni profiles are 
apparent, but only extending to a blue velocity of 2780 km s$^{-1}$, much 
smaller than the 6000 km s$^{-1}$ during the 1990 outburst \citep{S90}. 
These absorptions have been interpreted as jet ejections
in this binary which contains an M4.5 giant and a likely magnetic white
dwarf \citep{T92}. This picture is substantiated by the detection of 
radio emission during the outburst on 2016 April 5 \citep{Lu16}.

\section{Systems with Strong HeII}

Five objects in Figure 1 show noticeable \ion{He}{2}: V677 And, 0333+33, 
1005+69, 1432+19 and 2112-06 and, as discussed in the previous section, all
show velocities and periods consistent with having a magnetic white
dwarf. Of these, V677 And, 1005+69 and 1432+19
have this high excitation line stronger than H$\beta$ at some portion of
their orbits. The long period system 1005+69 is likely an IP, whereas the
other two are likely polars. Of the two weaker lined systems,  2112-06 is a 
known eclipsing polar and 0333+33 is likely a polar as well.

\section{Comments on Remaining Spectra}

\subsection{Unusual States}

Besides the symbiotic V694 Mon at outburst that was described in
the previous section, and V380 Oph in a low state discussed in section 3.2,
there are two other systems shown in Figure 1 that were
observed due to notifications that they were in unusual states. 
These are KR Aur that was coming out of a low state, and LS Peg that was
in one of its fainter observed states.

KR Aur is a novalike variable with a period of 3.9 hr that varies between high 
and low states of 11-18th mag with most of the time between mag 12-14 \citep{S83}. The AAVSO
lightcurves show that it was at 16.8 in 2017 Jan and about 15.7 in March
when the APO spectrum was obtained. Shafter obtained a spectrum when KR Aur
was at V=15.5 in 1982 Jan-Feb. Our specrum in Figure 1a is 2-3 times
brighter than that shown in \citet{S83} and is considerably
bluer. This is unexpected, as the mass transfer rate and the disk emission is
thought to be lower at low states. It is likely that Shafter's data were
also obtained during the rise to a high state, as a minimum at 17 mag was 
reported at the end of 1981 Dec \citep{P82}. Thus, it seems that the optical
magnitude is not sufficient to determine the state of accretion and the disk
contribution.

LS Peg is another novalike system that spends most of its time at a high
state (V=12) with a high mass transfer rate and high velocity emission
line wings \citep{G92,T99}, with occasional
drops to a low state near 14. Because of
some indications of a period near 20 min, it has been postulated as an IP, and
X-ray spectra concur with this classification even though no stable spin
period was detected \citep{R08}. The AAVSO light curves show LS Peg
was at 13.4 in 2015 July and 12.5 in Aug so it was coming out of a low state
when our spectrum was obtained at the end of June. Our spectrum in Figure 1b
compared to the Taylor et al. (1999) data (obtained in 1996-1997 when LS Peg
was in its high state) is 2-3 times fainter in the continuum
and line fluxes. 

\subsection{Typical Dwarf Novae}

About half of the remaining systems in Figure 1 have spectra that look
like typical dwarf novae with broad Balmer emission lines and a flat
Balmer decrement. These objects include
0033+38, 0150+33, 0206+20, 0359+17, 0422+33, 1853+42, 2246+06 and 2342+34.
0359+17 shows moderate to strong doubled lines, indicative of moderate
to high inclination. A couple of others show a narrow component superposed
which is likely a hot spot: 0206+20, 0422+33, 2319+33.

\subsection{Weak or Narrow Lines}

Four objects in Figure 1 have weaker lines than normal for a typical CV.
These objects are 0309+26, 1055+66, 1325-08 and 2059-09. They may have
been observed before they returned to their quiescent state. There are
also 3 systems with very narrow lines: 1626+33, 2200+25 and 2319+08.
The first object (1626+33) was identified as a potential
polar in the SDSS due to its very strong \ion{He}{2} line
\citep{S04}. Our
three spectra obtained in 2015 June are about a factor of 5 fainter than
the SDSS spectrum and have \ion{He}{2} weak to non-existent (Figure 1b).
It is likely this object was in a low state during our observations and
the rapidly changing strength of the \ion{He}{2} line during the 3
exposures indicates that a polar classification is more likely than an IP.
The second object (2200+25) has had 4 outbursts in the CRTS database and
is listed as a faint ($g$=20.9) blue object in SDSS DR12. While the continuum
is very faint, the Balmer emission lines are very strong, similar to V380 Oph
in its low state.
The third object 
2319+08 was observed as part of a list of ellipsoidal variables with periods
below 0.22 d and blue ${\it GALEX}$ colors that could be hidden CVs \citep{D16}.It was
listed with a period of 0.107760d (in the period gap). Our spectra in
2016 Nov and Dec show prominent TiO bands in the red and narrow Balmer 
emission so this is likely an irradiated M star system which could be a 
pre-CV. 

\section{Conclusions}

Our followup spectra have shown several systems with interesting properties
that merit further detailed study. Three systems (V677 And, 0333+33 and 2112-06)
are likely polars since they show the presence of \ion{He}{2} and high velocity 
amplitudes throughout their orbital periods. In addition, 1626+33 appears
much fainter in its continuum and \ion{He}{2} line strength than during the 
SDSS spectrum and is likely a polar that had gone into a low state. These 
objects merit circular polarimetry to 
confirm a high magnetic field strength for the white dwarfs. The system 1005+69
appears to be an IP and requires high speed photometry to search for a spin
period, while 1432+19 is likely an eclipsing SW Sex star with a period
below the gap. Photometry of the latter can confirm this short period and
determine an inclination from the eclipse. Two dwarf novae (0648+06 and 0853+48)
appear to have periods in the period gap, adding to the small number of
systems within this range. Spectra of three known novalike systems (V380 Oph, 
KR Aur and LS Peg) observed during low states are presented, while the
symbiotic V694 Mon was caught during an outburst, showing P Cygni profiles
indicative of jet ejections similar to the 1990 outburst. The information
gathered for these 38 objects over a span of almost 4 years shows the
large amount of effort that will be needed to sort out object types in
the forthcoming deluge of transients that will appear in ZTF and LSST.

\acknowledgments

PS and DSG aknowledge support from NSF grant AST-1514737. Z.D. acknowledges 
support from CAS Light of West
China Program and the Science Foundation of Yunnan Province (No.
2016FB007).
The students in Astro 497, 499 are acknowledged for their
help in obtaining spectra on the nights of 2016 April 21 (David Bordenave, 
Nicholas Huntley, Tessa Wilkinson) and 2017 March 5 (Ellis Avallone).

\clearpage
\startlongtable
\begin{deluxetable}{lllcccll}
\tabletypesize{\scriptsize}
\tablewidth{0pt}
\tablecaption{Summary of Spectroscopic Observations}
\tablehead{
\colhead{UT Date} & \colhead{Coords} & \colhead{Type\tablenotemark{a}} & 
\colhead{Source} & \colhead{Obs} 
& \colhead{UT start} & \colhead{Exp(sec)} & \colhead{State}}
\startdata
2016 Nov 23 & 003304+380105 & DN & CRTS,SDSS & APO & 04:41 & 900 & quies \\
2015 Sep 13 & 003827+250925 & DN & SDSS & APO & 05:00 & 600x10 & quies \\
2014 Sep 01 & 011614+092216 & DN & CRTS,SDSS & APO & 08:00 & 600x19 & quies \\
2016 Feb 11 & 011614+092216 & DN & CRTS,SDSS & APO & 01:41 & 900 & quies \\
2016 Nov 23 & 011614+092216 & DN & CRTS,SDSS & APO & 05:30 & 900 & quies \\
2016 Dec 05 & 011614+092216 & DN & CRTS,SDSS & APO & 05:02 & 900x6 & quies \\
2014 Aug 24 & 015052+332622 & DN & CRTS,SDSS & KPNO & 11:15  & 900 & quies \\
2014 Aug 24 & 020633+205708 & DN & CRTS,SDSS & KPNO & 10:50  & 1700 & quies \\
2014 Dec 15 & 030930+263804 & DN & CRTS,SDSS & APO & 01:29 & 600,900 & quies \\
2015 Dec 18 & 033357+332152 & P & CRTS & APO & 02:05 & 600,900x6 & quies \\
2014 Aug 23 & 035906+175034 & DN & CRTS,SDSS & KPNO & 11:16 & 1050x2 & quies \\
2014 Aug 23 & 041139+232220 & DN & CRTS & KPNO & 10:29  & 1200x2 & quies \\
2014 Dec 15 & 041139+232220 & DN & CRTS & APO & 02:01 & 900x8 & quies \\
2014 Aug 24 & 042218+334215 & DN & CRTS,SDSS & KPNO & 11:35 & 1050 & quies \\
2017 Mar 05 & 050124+203818 & DN & CRTS & APO & 02:47 & 900x7 & quies \\
2017 Mar 05 & 061544+283509 & NL & KR Aur & APO & 05:01 & 600 & mid \\
2016 Feb 11 & 062658+242907 & DN & USNO & APO & 05:36 & 600x2 & out \\
2014 Dec 15 & 064833+065624 & DN & PNV & APO & 04:28 & 900x4,600x8 & quies \\
2016 Mar 15 & 072551-074408 & S & V694 Mon & APO & 02:19 & 300 & out \\
2016 Mar 15 & 085333+484847 & DN & SDSS & APO & 02:53 & 600x6 & quies \\
2016 Apr 12 & 085333+484847 & DN & SDSS & APO & 02:26 & 600x5,900x4 & quies \\
2017 Mar 05 & 092919+622346 & DN & SDSS & APO & 05:21 & 600x7 & quies \\ 
2014 Mar 23 & 100517+694136 & IP? & SDSS & APO & 05:20 & 900x5 & quies \\
2014 May 04 & 100517+694136 & IP? & SDSS & APO & 05:08 & 900x5 & quies \\
2014 May 06 & 100517+694136 & IP? & SDSS & APO & 02:46 & 900x13 & quies \\
2015 Jun 22 & 105504+681208 & DN & vsnet & APO & 03:12 & 300,600 & quies \\
2016 May 03 & 124539-073706 & DN & CRTS,SDSS & APO & 02:50  & 900x7 & quies \\
2015 Jun 22 & 132530-082009 & DN & CRTS & APO & 05:02 & 900x2 & quies \\
2016 Apr 12 & 143210+191403 & NL & SDSS & APO & 04:47 & 600x4 & quies \\
2016 May 03 & 143210+191403 & NL & SDSS & APO & 05:03 & 600x10 & quies \\
2014 Jun 21 & 154532+010226 & DN & ASASSN,SDSS & APO & 04:57 & 300x12 & out \\
2015 Jun 22 & 162608+332828 & P? & SDSS & APO & 06:04 & 600x3 & quies \\
2014 Jun 09 & 164707+622451 & DN & CRTS,SDSS & KPNO & 03:57 & 1200 & out \\
2015 Sep 13 & 175014+060529 & NL & V380 Oph & APO & 02:00 & 900x10 & low \\
2014 Nov 21 & 185329+420343 & DN & ASASSN & APO & 01:02 & 600 & out \\
2014 Dec 15 & 185329+420343 & DN & ASASSN & APO & 00:53 & 600,900 & quies \\
2016 Oct 25 & 205933-091616 & DN & PTF & APO & 01:59 & 600x2 & quies \\
2016 Nov 23 & 211203-061638 & P & ASASSN,SDSS & APO & 01:09 & 600x8 & quies \\
2015 Jun 22 & 215158+140653 & NL & LS Peg & APO & 06:52 & 300x2 & low \\
2014 Aug 27 & 220019+254551 & DN & CRTS,SDSS & KPNO & 09:52 & 1200x2 & quies \\
2016 Oct 25 & 220031+033430 & DN & CRTS & APO & 02:30 & 600,900x4 & quies \\
2016 Nov 23 & 220031+033430 & DN & CRTS & APO & 02:55 & 600,900x5 & quies \\
2014 Aug 24 & 224648+065635 & DN & CRTS,SDSS & KPNO & 10:25 & 1200 & quies \\
2016 Nov 23 & 231910+082832 & DN & CRTS,SDSS & APO & 05:10 & 600 & quies \\
2016 Dec 05 & 231910+082832 & DN & CRTS,SDSS & APO & 01:54 & 600x4 & quies \\
2016 Dec 05 & 231910+331540 & DN & CRTS,SDSS & APO & 02:49 & 600x11 & quies \\
2015 Oct 17 & 233423+391423 & P & CRTS, V677 And & APO & 01:22 & 900x12 & quies \\
2014 Aug 27 & 234242+341331 & DN & CRTS,SDSS & KPNO & 10:38 & 1200 & quies \\
2015 Nov 18 & 235052+285859 & DN & Gaia & APO & 01:11 & 900x4,692 & quies \\
2015 Dec 18 & 235052+285859 & DN & Gaia & APO & 04:04 & 900x5 & quies \\
\enddata
\tablenotetext{a}{Provisional type of Dwarf Nova (DN), Polar (P), Intermediate
Polar (IP), Novalike (NL), Symbiotic (S)}
\end{deluxetable}
\clearpage

\startlongtable
\begin{deluxetable}{lccc}
\tabletypesize{\scriptsize}
\tablewidth{0pt}
\tablecaption{Equivalent Widths of Emission Lines (\AA)}
\tablehead{
\colhead{Object} & \colhead{H$\beta$} & \colhead{H$\alpha$} & \colhead{HeII} }
\startdata
0033+38 & 59 & 59 & ... \\
0038+25 & 53 & ... & ... \\
0116+09 & 33 & 77 & ... \\
0150+33 & 78 & ... & 10 \\
0206+20 & 96 & ... & ... \\
0309+26 & 4 & 28 & ... \\
0333+33 & 5 & 16 & 3 \\
0359+17 & 34 & ... & ... \\
0411+23 & 24 & 30 & ... \\
0422+33 & 25 & ... & ... \\
0501+20 & 27 & 37 & ... \\
KR Aur & 65 & 113 & 9 \\
0648+06 & 96 & 94 & ... \\
V694 Mon & 14 & 81 & 0.2 \\
0853+48 & 39 & 60 & ... \\
0929+62 & 10 & 11 & ... \\
1005+69 & 78 & 64 & 48 \\
1055+68 & 9 & 21 & ... \\
1245-07 & 57 & 74 & ... \\
1325-08 & 12 & 25 & ... \\
1432+19 & 17 & 24 & 29 \\
1626+33 & 37 & 51 & ... \\
V380 Oph & 22 & ... & ... \\
1853+42 & 27 & 54 & ... \\
2059-09 & 9 & 22 & 11 \\
2112-06 & 11 & 11 & 2 \\
LS Peg & 9 & 22 & 1 \\
2200+25 & 68 & ... & ... \\
2200+03 & 37 & 34 & ... \\
2246+06 & 61 & ... & ... \\
2319+08 & 4 & 6 & ... \\
2319+33 & 58 & 72 & ... \\
V677 And & 21 & 14 & 13 \\
2342+34 & 59 & ... & 6 \\
2350+28 & 5 & 11 & ... \\
\enddata
\end{deluxetable}

\clearpage
\begin{deluxetable}{llcccc}
\tabletypesize{\scriptsize}
\tablewidth{0pt}
\tablecaption{Radial Velocity Fits}
\tablehead{
\colhead{Object} & \colhead{Line} & \colhead{P(min)} & 
\colhead{$\gamma$(km s$^{-1}$)} & \colhead{K (km s$^{-1}$)} &
\colhead{$\sigma$ (km s$^{-1}$)} } 
\startdata
V677 And & H$\beta$ & 104 & -28.6$\pm$0.1 & 402$\pm$7 & 16 \\
V677 And & HeII & 105 & 8.7$\pm$0.2 & 400$\pm$24 & 55 \\
V380 Oph & H$\beta$ & 222 & 9.6$\pm$0.9 & 71$\pm$3 & 6 \\
0038+25 & H$\beta$ & 95 & 87.8$\pm$2.4 & 95$\pm$13 & 27 \\
0116+09 & H$\alpha$ & 94 & 8.5$\pm$0.9 & 76$\pm$1 & 15 \\
0333+33 & H$\alpha$ & 110.5\tablenotemark{a} & 88$\pm$1 & 218$\pm$31 & 54 \\
0333+33 & H$\beta$ & 110.5\tablenotemark{a} & 85$\pm$2 & 243$\pm$45 & 70 \\
0411+23 & H$\alpha$ & 202 & 41.5$\pm$1.3 & 34$\pm$4 & 5 \\ 
0501+20 & H$\alpha$ & 108 & 45$\pm$1 & 84$\pm$5 & 8 \\
0501+20 & H$\beta$ & 108\tablenotemark{a} & 37$\pm$3 & 51$\pm$19 & 33 \\
0648+06 & H$\alpha$ & 145 & -26.3$\pm$0.3 & 71$\pm$3 & 7 \\
0648+06 & H$\beta$ & 138 & -35.9$\pm$0.5 & 67$\pm$4 & 9 \\
0853+48 & H$\alpha$ & 142 & -84$\pm$2 & 96$\pm$25 & 51 \\
0853+48 & H$\beta$ & 142 & -182$\pm$3 & 135$\pm$27 & 54 \\
1005+69 & H$\alpha$ & 220 & 53$\pm$13 & 211$\pm$19 & 46 \\
1005+69 & H$\beta$ & 220 & 76$\pm$14 & 248$\pm$19 & 48 \\
1005+69 & HeII & 220 & 87$\pm$15 & 218$\pm$21 & 52  \\
1245-07 & H$\alpha$ & 118 & -3.3$\pm$1.5 & 60$\pm$18 & 27 \\
1432+19 & HeII & 103 & -109$\pm$3 & 98$\pm$14 & 25 \\
1432+19 & H$\beta$ & 103\tablenotemark{a} & -149$\pm$5 & 96$\pm$21 & 43 \\ 
2112-06 & H$\alpha$ & 95.7\tablenotemark{a} & 71$\pm$7 & 357$\pm$39 & 59 \\
2319+33 & H$\alpha$ & 171\tablenotemark{a} & 21$\pm$2 & 60$\pm$5 & 11 \\
2319+33 & H$\beta$ & 171 & -12$\pm$3 & 53$\pm$10 & 16 \\
2350+28 & H$\alpha$ & 77 & -1.3$\pm$0.4 & 94$\pm$9 & 11 \\ 
\enddata
\tablenotetext{a}{Period fixed at this value}
\end{deluxetable}

\clearpage

\begin{figure}
\figurenum {1a}
\includegraphics[width=6in]{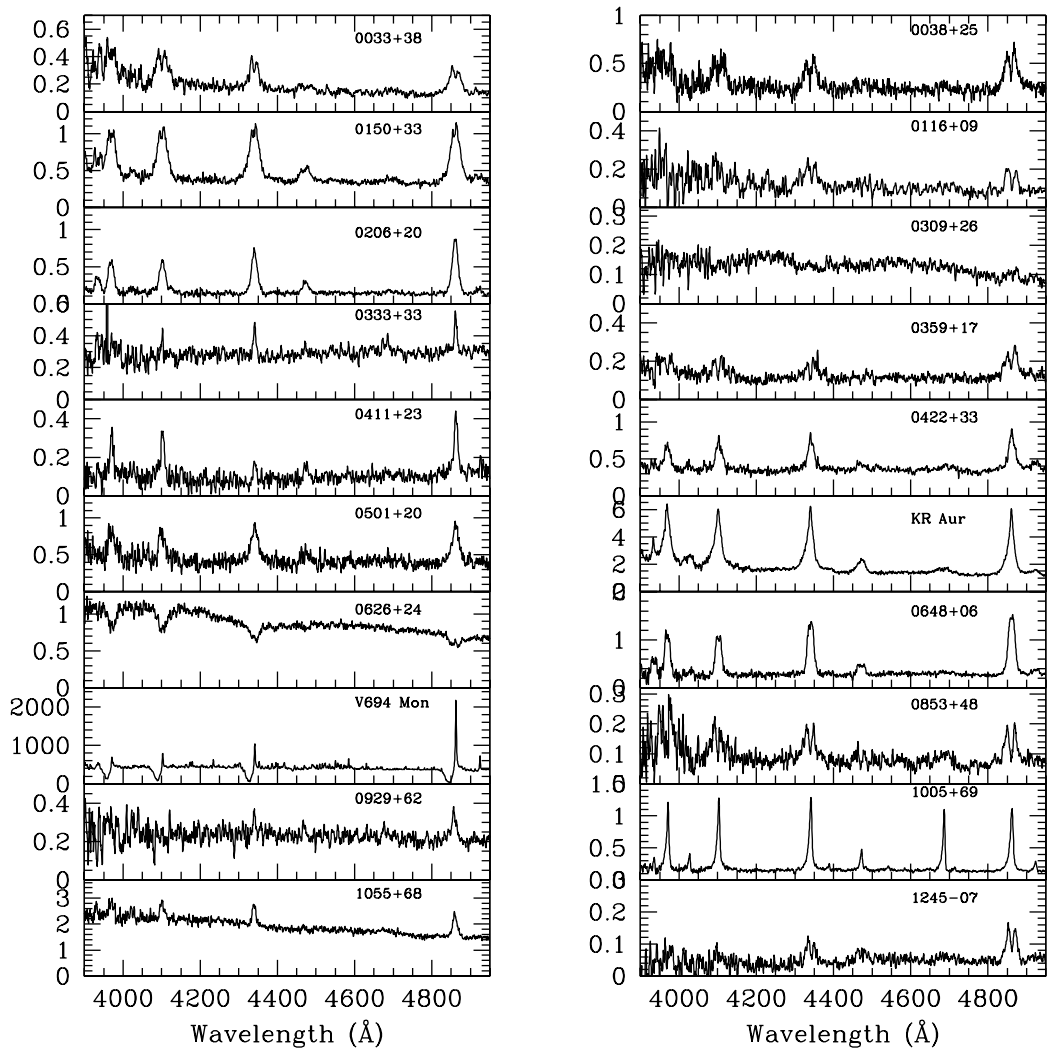}
\caption{Blue region spectra of sources listed in Table 1. Vertical axes are
F$_{\lambda}$ in units of 10$^{-15}$ ergs cm${-2}$ s$^{-1}$ \AA$^{-1}$.
Objects are labelled with first digits of RA and Dec as given in Table 1.}
\end{figure}

\clearpage

\begin{figure}
\figurenum {1b}
\includegraphics[width=6in]{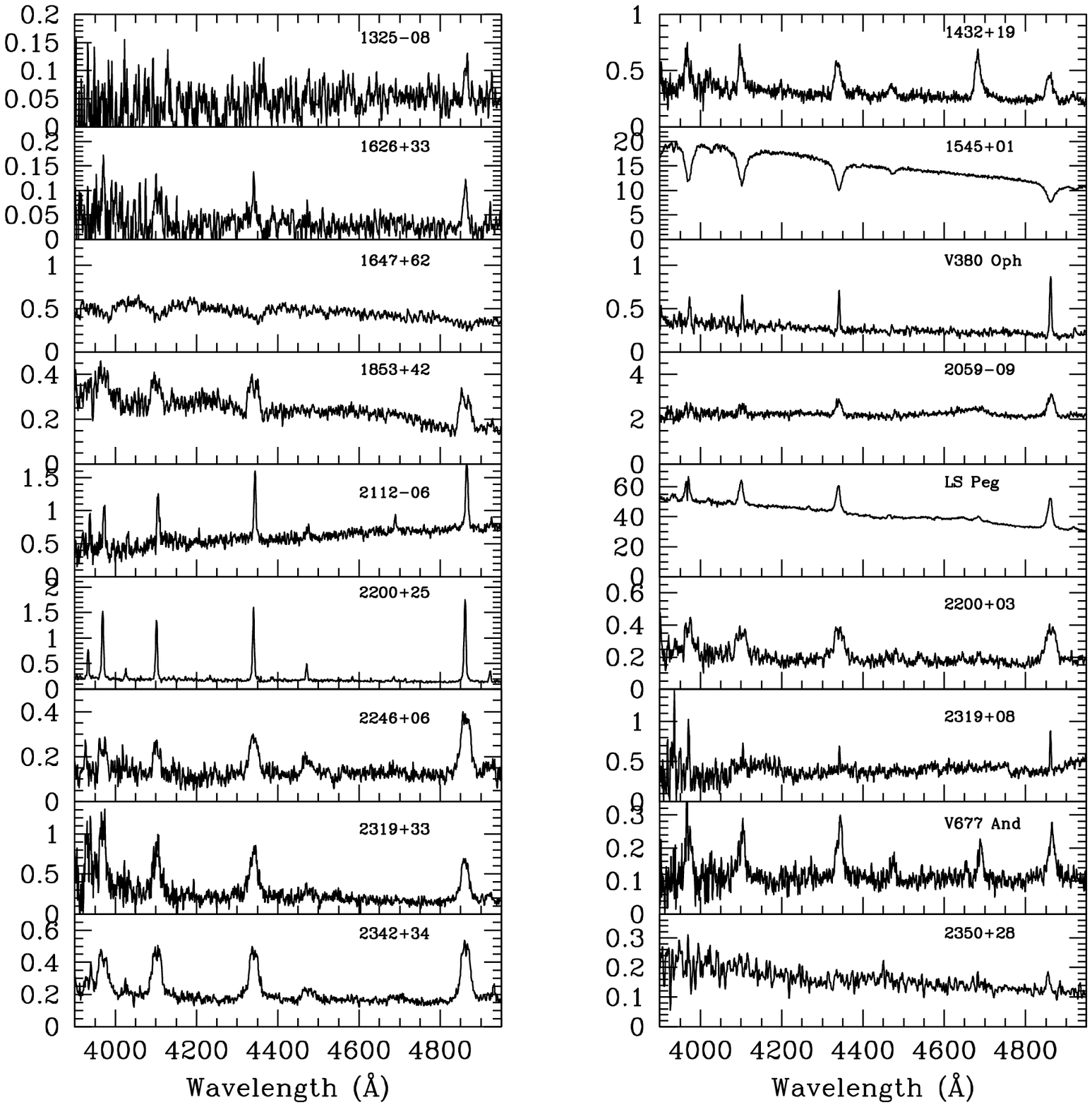}
\caption{Figure 1 continued.}
\end{figure}

\clearpage

\begin{figure}
\figurenum {2}
\includegraphics[width=6in]{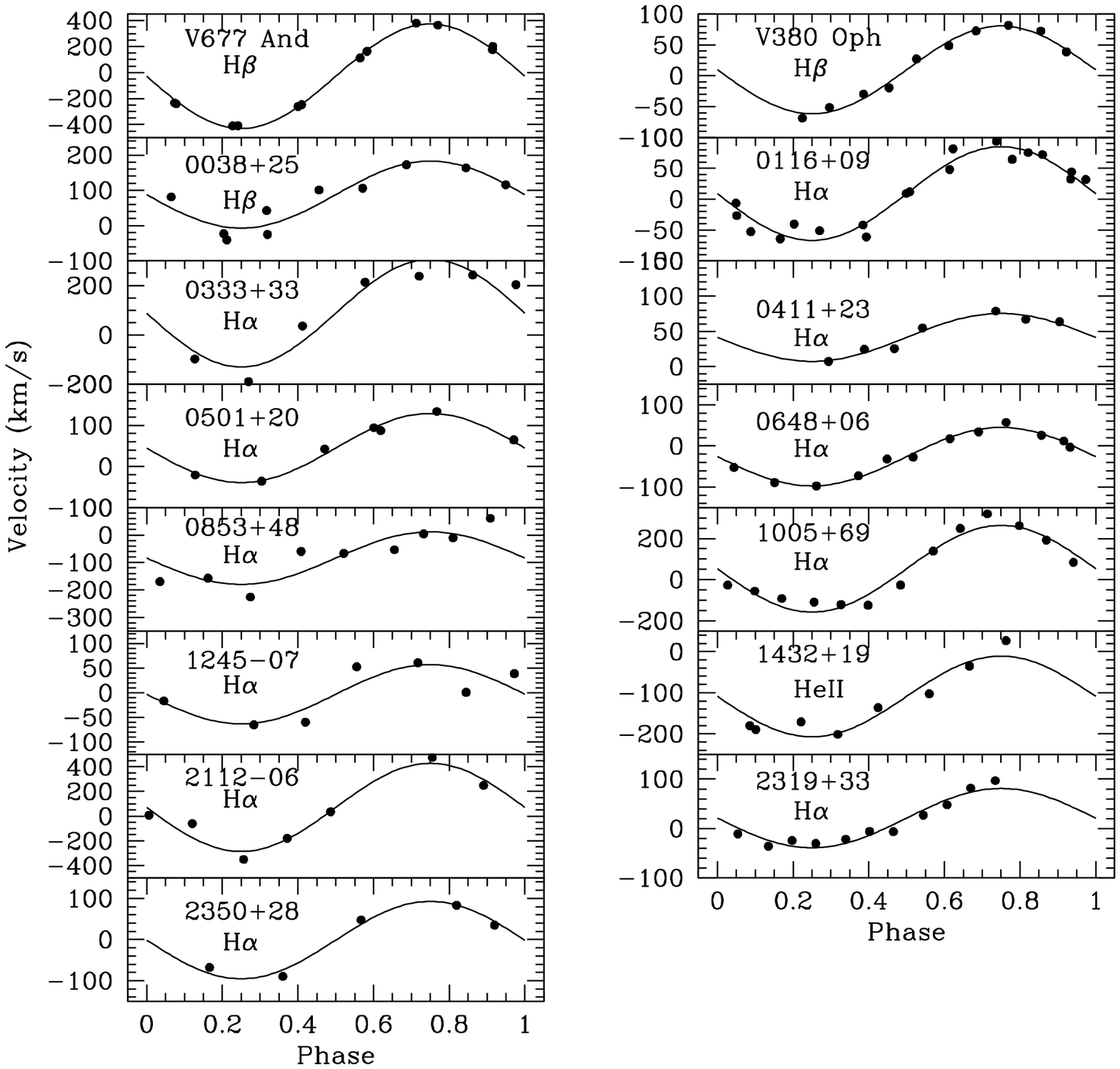}
\caption{Radial velocity curves of the 15 systems with time-resolved
spectra along with the best-fit sine curves with parameters as listed in Table 
2.}
\end{figure}

\clearpage
\begin{figure}
\figurenum {3}
\includegraphics[width=6in]{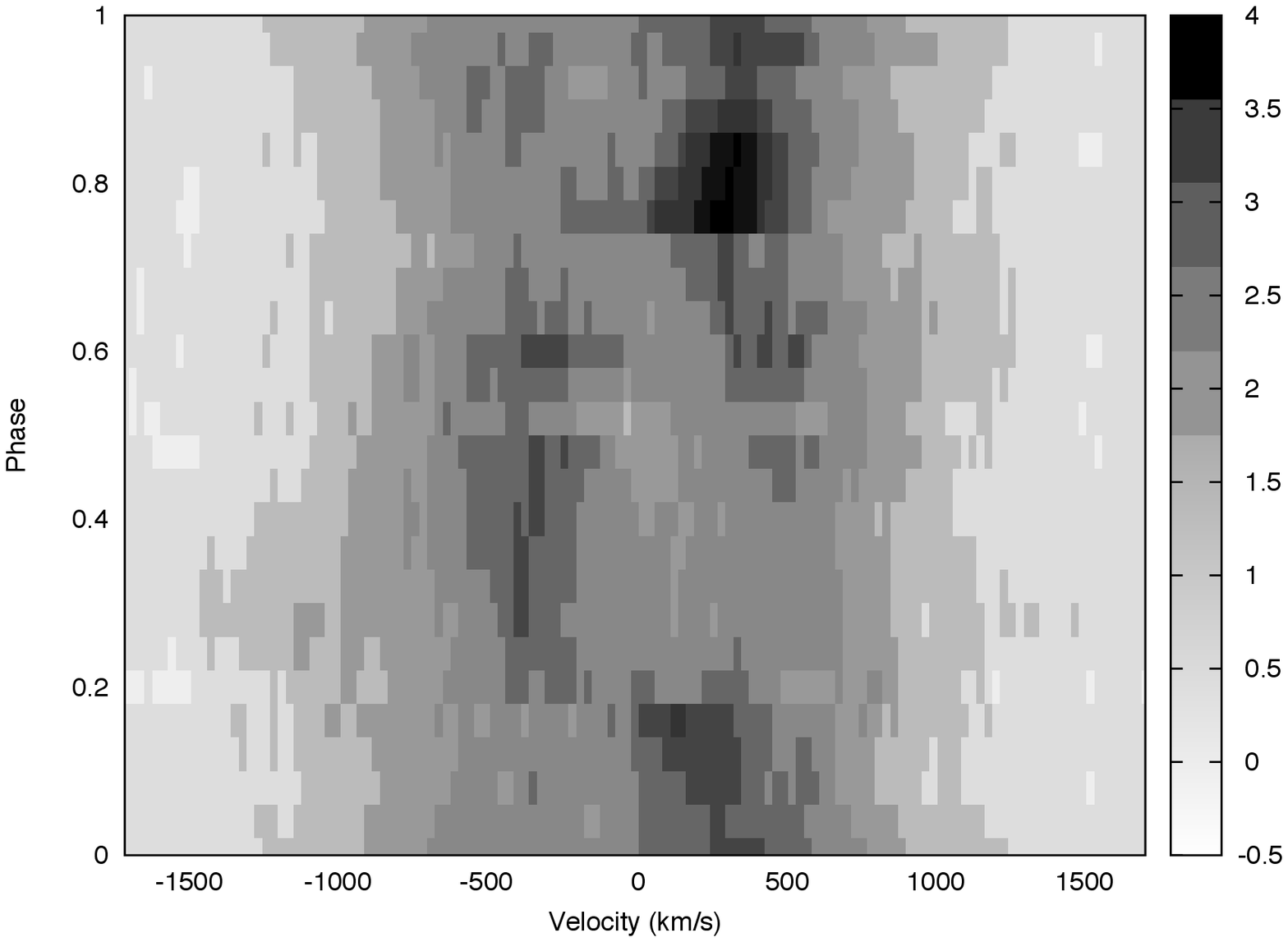}
\caption{Trailed spectra of 0116+09 from 2014 September 01 spectra.}
\end{figure}

\clearpage
\begin{figure}
\figurenum {4}
\includegraphics[width=6in]{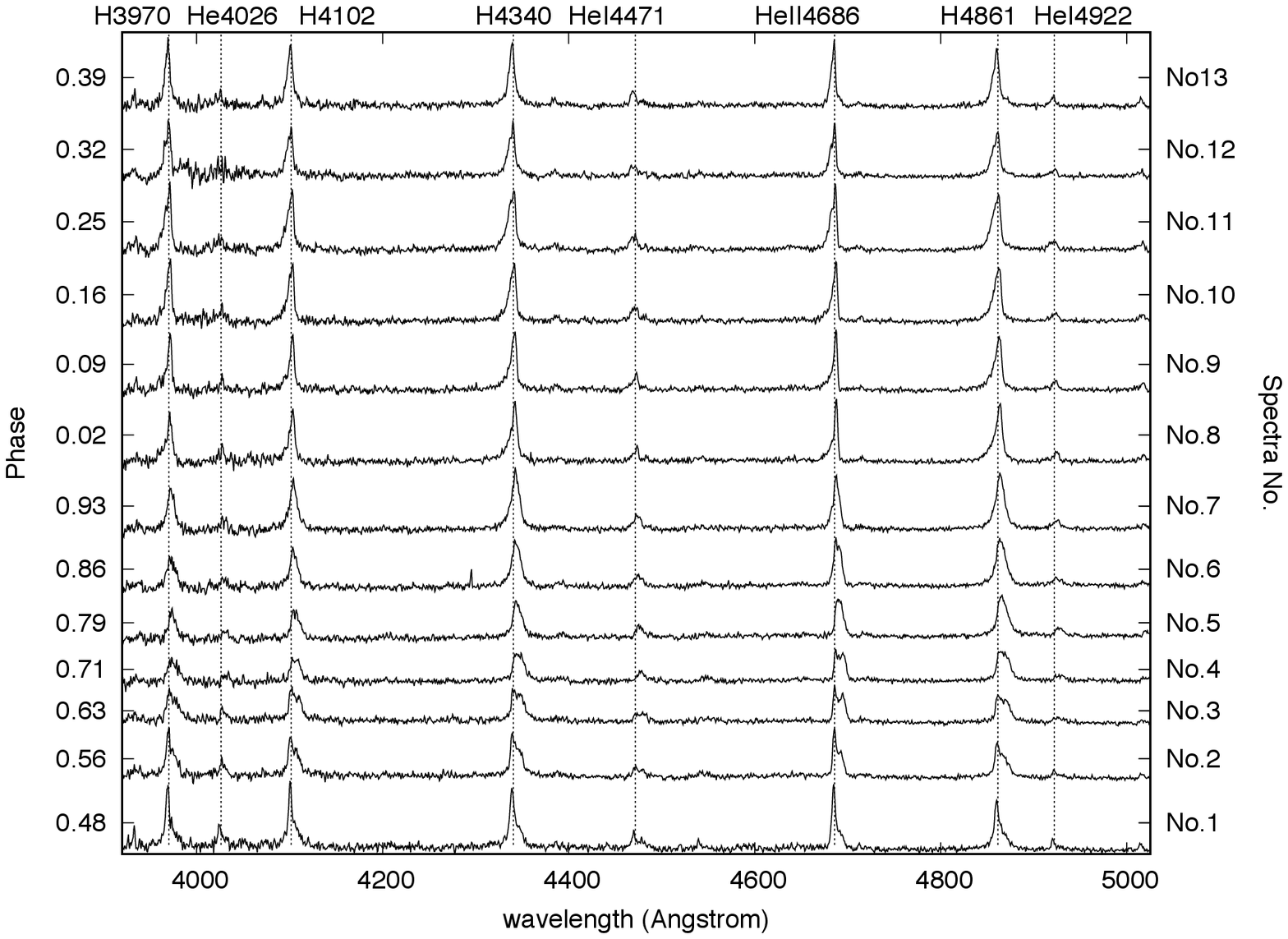}
\caption{The 2014 May 6 spectra of 1005+69 showing the changing line
shapes throughout its orbit.}
\end{figure}


\begin{thebibliography}{}

\bibitem[Bond(1979)]{B79} Bond, H. E. 1979, IAU Coll. 53, 495

\bibitem[Breedt et al.(2014)]{B14} Breedt, E., G\"ansicke, B. T., Drake, A. J. et al. 2014, \mnras, 443, 3174

\bibitem[Cook(2015)]{C15} Cook, L. 2015, vsnet-alert 19078

\bibitem[Coppejans et al.(2014)]{C14} Coppejans, D. L., Woudt, P. A., Warner, B. et al. 2014, \mnras, 437, 510

\bibitem[Denisenko(2016)]{D16} Denisenko, D. 2016, vsnet-alert 19859

\bibitem[Drake et al.(2009)]{D09} Drake, A. J., Djorgovski, S. G., Mahabal, A. et al. 2009, \apj, 696, 870

\bibitem[Drake et al.(2014)]{D14} Drake, A. J., G\"ansicke, B. T., Djorgovski, S. G., Wils, P. et al. 2014, \mnras, 441, 1186

\bibitem[G\"ansicke et al.(2009)]{G09} G\"ansicke, B. T., Dillon, M., Southworth, J. et al. 2009, \mnras, 397, 2170

\bibitem[Garnavich \& Szkody(1992)]{G92} Garnavich, P. \& Szkody, P. 1992, JAAVSO, 21, 81

\bibitem[Hoard et al.(2003)]{H03} Hoard, D. W., Szkody, P., Froning, C.S., Long, K. S., Knigge, C. 2003, \aj, 126, 2473

\bibitem[Lipunov et al.(2010)]{L10} Lipunov, V. M., Kornilov, V., Gorbovskoy, E. et al. 2010, AdAst, 2010, 349171

\bibitem[Littlefield et al.(2015)]{L15} Littlefield, C., de Miguel, E., Cook, L. 2015, ATel 8368

\bibitem[Littlefield et al.(2016)]{L16} Littlefield, C. Cook. L. M., Bersier, D. et al. 2016, ATel 9764

\bibitem[Lucy et al.(2016)]{Lu16} Lucy, A. B., Weston, J. H. S., Sokoloski, J. L. 2016, Atel 8957

\bibitem[Maehara(2014)]{M14} Maehara, H. 2014 vsnet-alert 18004

\bibitem[Mahabal et al.(2008)]{M08} Mahabal, A., Drake, A. J., Djorgovski, S. G. et al. 2008, ATel 1741

\bibitem[Munari et al.(2016)]{M16} Munari, U., Dallaporta, S., Castellani, F. et al. 2016, New Astr., 49, 43

\bibitem[Pojmanski(1997)]{P97} Pojmanski, G. 1997, AcA, 47, 467

\bibitem[Popov(1982)]{P82} Popov, V. 1982, IBVS 2095

\bibitem[Quimby et al.(2008)]{Q08} Quimby, R., Rau, A., Ofek, E. et al. 2008, ATel 1750

\bibitem[Ramsay et al.(2008)]{R08} Ramsay, G., Wheatley, P. J., Norton, A. J. et al. 2008, \mnras, 387, 1157

\bibitem[Rodriguez-Gil et al.(2007)]{R07} Rodriguez-Gil, P., Schmidtobreick, L., G\"ansicke, B. T. 2007, \mnras, 374, 1359

\bibitem[Schmeer(2014)]{S14} Schmeer, P. 2014, vsnet-alerts 18001,18003

\bibitem[Shafter(1983)]{S83} Shafter, A. W. 1983, \apj, 267, 222

\bibitem[Shafter(1985)]{S85} Shafter, A. W. 1985, \aj, 90, 643

\bibitem[Shappee et al.(2014)]{Sh14} Shappee, B. J., Prieto, J. L., Grupe, D. et al. 2014,\apj, 788, 48

\bibitem[Shugarov et al.(2005)]{S05} Shugarov, S. Y., Katysheva, N. A., Seregina, T. M., Volkov, I. M., Kroll, P. 2005, ASPCS, 330, 495


\bibitem[Szkody et al.(2004)]{S04} Szkody, P., Henden, A., Fraser, P., Silvestri, N. et al. 2004, \aj, 128, 1882

\bibitem[Szkody et al.(2011)]{S11} Szkody, P., Anderson, S. F., Brooks, K. et al. 2011, \aj, 142, 181

\bibitem[Szkody et al.(2014)]{Sz14} Szkody, P., Everett, M. E., Howell, S. B., Landolt, A. U., Bond, H. E. et al. 2014, \aj, 148, 63

\bibitem[Szkody et al.(1990)]{S90} Szkody, P., Mateo, M. \& Schmeer, P. 1990, IAUC. 4987

\bibitem[Taylor et al.(1999)]{T99} Taylor, C., Thorstensen, J., Patterson, J. P., 1999, \pasp, 111, 184

\bibitem[Thorstensen \& Skinner(2012)]{T12} Thorstensen, J. R. \& Skinner,, J. N. 2012, \aj, 144, 81

\bibitem[Thorstensen et al.(2016)]{T16} Thorstensen, J. R., Alper, E. H., Weil, K. E. 2016, \aj, 152, 226

\bibitem[Tomov et al.(1992)]{T92} Tomov, T., Zamanov, D., Kolev, L. et al. 1992, \mnras, 258, 23 

\bibitem[Warner(1995)]{W95} Warner, B. 1995, Cataclysmic Variable Stars (Cambridge: Cambridge University Press)

\bibitem[Wei et al.(2013)]{W13} Wei, P., Luo, A., Li, Y. et al. 2013, \mnras, 431, 1800

\bibitem[Wevers et al.(2014)]{W14} Wevers, T., Jonker, P. G., van Velzen, S. et al. 2014, ATel 6788

\bibitem[York et al.(2000)]{Y00} York, D. G., Adelman, J., Anderson, J. E. et al. 2000, \aj, 120, 1579

\end{thebibliography}
\end{document}